**The design and statistical aspects of VIETNARMS: a strategic post-licensing trial of multiple oral direct acting antiviral Hepatitis C treatment strategies in Vietnam**

Leanne McCabe[1], Ian R. White[1], Nguyen Van Vinh Chau[2], Eleanor Barnes[3], Sarah L. Pett[1], Graham S. Cooke[4], A. Sarah Walker[1] on behalf of SEARCH investigators

[1] Medical Research Council Clinical Trials Unit at University College London, London, UK

[2] Hospital for Tropical Diseases, Ho Chi Minh City, Vietnam

[3] Oxford University, Oxford, UK

[4] Imperial College London, London, UK

**ABSTRACT**

**Background**

Achieving hepatitis C elimination is hampered by the costs of direct acting antiviral treatment and the need to treat hard-to-reach populations. Treatment access could be widened by shortening or simplifying treatment, but limited research means it is unclear which treatments or strategies could achieve sufficiently high cure rates to be acceptable, particularly in settings with a high prevalence of less common viral genotypes. We present the statistical aspects of a multi-arm trial designed to test multiple strategies simultaneously with a monitoring mechanism to detect and stop those with unacceptably low cure rates quickly.

**Methods**

The VIETNARMS trial will factorially randomise patients to two drug regimens, three treatment shortening strategies or control, and adjunctive ribavirin or no adjunctive ribavirin with shortening strategies (14 randomised groups). We will use Bayesian monitoring at interim analyses to detect and stop recruitment into unsuccessful strategies, defined as a >0.95 posterior probability of the true cure rate being <90%. Final comparisons will be non-inferiority for regimen and strategy comparisons and superiority for the ribavirin comparison. Here, we tested the operating characteristics of the stopping guideline, planned the timing of the interim analyses and explored power at the final analysis.

**Results**

A beta(4.5, 0.5) prior for the true cure rate produces <0.05 probability of incorrectly stopping a group with true cure rate >90%. Groups with very low cure rates (<60%) are very likely (>0.9 probability) to stop after ~25% patients are recruited. Groups with moderately low cure rates (80%) are likely to stop (0.7 probability) before the end of recruitment. Interim analyses 7, 10, 13 and 18 months after recruitment commences provide good probabilities of stopping inferior groups. For an overall true cure rate of 95%, power is >90% to detect non-inferiority in the regimen and strategy comparisons using 5% and 10% margins respectively, regardless of the control cure rate, and to detect a 5% absolute difference in the ribavirin comparison.

**Conclusions**

The operating characteristics of the stopping guideline are appropriate and interim analyses can be timed to detect failing groups at various stages. Our design is therefore suitable for selecting treatment shortening or simplifying strategies.

**Trial Registration**

ISRCTN registry: ISRCTN61522291. Registered on 4$^{th}$ October 2019.

**KEYWORDS**

Bayesian methods, Clinical trial, Hepatitis C, Interim analyses, Multi-arm, Trial design

**BACKGROUND**

Oral direct acting antivirals (DAAs) have transformed the treatment of hepatitis C (HCV). Compared to historical injectable interferon-based treatment of 6-12 months, they are more effective, better tolerated and offer shorter durations of therapy (8-12 weeks). However, access to treatment is still limited by costs, particularly in low-income countries where the patient pays at least in part. Further, achieving the WHO target for elimination of viral hepatitis as a public health threat by 2030 (1) will require curing hard-to-reach populations including the homeless, drug users and prisoners who still find adherence challenging. Strategies designed to reduce drug exposure while still achieving HCV cure, could further widen treatment access, but there has been limited research to date into shortening and simplifying HCV treatment with DAAs.

Previous very small studies assessing shortened DAA treatment have found higher cure rates when treatment length is guided by early response to treatment (2-4), or when adding adjunctive therapies, such as pegylated interferon (PEG-IFN) (5), to DAAs. Outside of HCV treatment, drug sparing strategies that allow intermittent dosing are widely used in the treatment of tuberculosis, allowing patients who may not comply with daily treatment to access supervised treatment. Although these approaches may also be successful in HCV, there are currently no studies assessing such a strategy with DAAs. There has also been little research into the use of DAAs in genotype 6, a strain that is most prevalent in Vietnam and surrounding countries (~50%), but uncommon elsewhere (<5%) (6). Several small studies have shown very high cure rates with DAAs in this genotype (7-10), as in other genotypes, but these have been mostly limited in regimen and to standard length courses. Finally, three pan-genotypic regimens (sofosbuvir/velpatasvir, sofosbuvir/daclatasvir, pibrentasvir/glecaprevir) are recommended by the most recent WHO guidelines (11), but there has been no direct randomised comparison of these regimens to date in any genotype.

As there is very little data to inform optimal ways to shorten HCV treatment, especially in genotype 6, it is possible that any proposed shortening strategy may fail. Therefore trial designs that include multiple different options whilst allowing for the early stopping of unsuccessful treatments in order to focus on more successful treatments are essential, both for trial efficiency and to protect patients. Two trial designs that incorporate both of these aspects are factorial trials and multi-arm multi-stage (MAMS) trials. Both designs allow for greater efficiency in trials by reducing the number of patients required and shortening the time needed to test multiple interventions compared to sequential trials of individual interventions (12, 13). However, as the timing and maximum number of interim analyses within MAMS trials have to be pre-specified (14) they may not be suitable for interventions where the effect on outcomes is unknown and additional analyses may have to be scheduled.

Data monitoring and stopping guidelines are most commonly framed within a frequentist framework with guidelines based on p-values or conditional power. When planning interim analyses care must be taken to control type I error, which can limit the ability to change the monitoring schedule to adapt to accumulating data, which may lead to delays in stopping unsuccessful treatments (if strict guidelines such as Haybittle-Peto (p<0.001) are used) or treatments being prematurely stopped. Monitoring can be improved by using a Bayesian approach, which allows for stopping guidelines that are based on directly interpretable probabilities and, as the prior is continuously updated, greater flexibility in the timing and numbers of interim analyses (15-18). Incorporating Bayesian monitoring within a multi-arm factorial trial allows for a flexible monitoring schedule to test multiple strategies and detect inferior ones quickly.

Here, we present the statistical aspects of the design of a multi-arm trial to be conducted in Vietnam that aims to find efficacious drug-sparing treatment strategies that will allow access to HCV treatment to be widened, with a particular focus on increasing evidence on treatment of genotype 6.

**METHODS**

**Trial design**

VIETNARMS is a parallel-group open-label factorial trial (ISRCTN61522291). 1092 patients will be factorially randomised: 1:1 to two different WHO recommended dual DAA regimens (sofosbuvir/velpatasvir vs sofosbuvir/daclatasvir); 1:2:2:2 to standard licenced 12-week treatment vs 4 weeks treatment with PEG-IFN+DAA vs 4-12 weeks response guided therapy (RGT) vs 12 weeks treatment using an induction/maintenance approach; and, if not randomised to standard 12-week treatment, 1:1 to adjunctive ribavirin vs no ribavirin for the duration of their DAA treatment (Figure 1). Randomisation will be stratified by genotype 6 vs all other genotypes. Patients randomised to the PEG-IFN strategy will receive DAAs for 4 weeks with weekly PEG-IFN for 4 weeks starting at day 7. Treatment length for those randomised to RGT will be determined by HCV viral load (VL) at day 7 and based on predicted viral kinetics (19): those with VL <lower limit of quantification (LLOQ) will receive 4 weeks of treatment, those with VL LLOQ-250 IU/ml will receive 8 weeks and all others will receive 12 weeks. Patients randomised to induction/maintenance will receive 12 weeks of treatment: 2 weeks of daily treatment (induction phase) followed by 10 weeks of 5 days treatment per week taking weekends off from the first weekend following their full 2 weeks treatment. Each strategy reduces DAA exposure and has other benefits, for example, compatibility with directly observed therapy programmes (as used for tuberculosis), but with other additional costs (Table 1). Within the trial, any patient not achieving cure with their first-line treatment will receive 12 weeks retreatment with the alternate drug regimen to the one they were originally randomised plus ribavirin.

**Primary endpoint**

The primary endpoint is Sustained Virologic Response (SVR12) i.e. virological cure on first-line therapy defined as plasma HCV VL <LLOQ 12 weeks after the end of first-line treatment (EOT+12) without prior failure. Failure is defined as either two consecutive HCV VL >LLOQ after two consecutive HCV VL <LLOQ with the latter confirmatory VL >2000 IU/ml, or two consecutive HCV VL

>1log10 increase above HCV nadir on treatment and >2000 IU/ml (either definition being met whilst on treatment or after finishing treatment during follow-up).

These failure criteria identify patients who, were they not to receive retreatment before EOT+12 weeks, would definitively have HCV VL >LLOQ at 12 weeks post-EOT and hence would be considered as failures in the primary endpoint. However, for ethical reasons, within the trial such patients will be offered retreatment as soon as they are definitively identified as having failed first-line treatment. Failure is defined using a higher threshold than the LLOQ because patients have been observed to achieve cure despite having low-level viraemia at EOT or shortly after, and so they do not need retreatment to achieve cure on first-line treatment. In practice, any participant with low-level viraemia <2000 IU/ml either cures or viral load rises above this level. This will be carefully reviewed by the independent Data Monitoring Committee (DMC). This is the same definition as used in the UK STOP-HCV-1 trial (ISRCTN37915093) and SEARCH-1 trial in Vietnam (ISRCTN17100273), where the DMC has similarly reviewed individual patient viral load trajectories.

**Monitoring**

For the strategies to be viable outside the trial, first-line cure rates need to be high (>90%). The design of the trial therefore allows for failing groups to be stopped early at any time and subsequent patients to be randomised to more successful groups. Individual performance of groups receiving shortening strategies will be monitored during recruitment by an independent DMC who will make decisions on whether a group should be stopped. Groups receiving standard 12-week treatment will not be monitored as this is the licenced duration with cure rates >90% (20, 21). Interim analyses will not be comparative as the aim of monitoring is not to find the best strategy, but to find any strategy that meets a minimum acceptable cure rate that may also be non-inferior to standard treatment as different strategies may benefit different patient populations.

Analyses of cure rates will follow the Bayesian paradigm to allow the probability of the true cure rate being below different thresholds to be calculated: recruitment into a group will stop if there is a >0.95 posterior probability of the true cure rate being <90% (Pr(true cure rate <0.9|x)>0.95 where x is the data currently observed). The primary monitoring is combined across genotypes; if the combined group reaches the stopping guideline, each genotype stratum will be tested separately and the DMC will have the discretion to stop only those strata reaching the stopping criteria. Differences in stopping groups across strata is only likely to occur when there are extreme differences in the cure rates between the strata, which is not expected, and so the operating characteristics of the trial are based on stopping combined strata only. If neither stratum reaches the stopping criteria despite the combined strata doing so, it will be at the discretion of the DMC whether to stop recruitment into

At interim analyses, there is greater uncertainty about the performance of the shortening strategies. Therefore, when determining the prior it was assumed that one strategy would fail completely such that all 4 groups receiving that strategy, of a total of 12 tested, would meet the stopping guideline. The mean of the prior was fixed at 0.9 and the effective sample size of the prior was varied until a distribution was found such that there was a ~0.33 probability of a cure rate <90%. The prior chosen was beta(4.5, 0.5) with mean 0.9, variance 0.015 and a 0.34 probability of a cure rate <90%. The relatively low precision of the prior will allow greater influence of the data in the posterior distribution. If the stopping guideline is met, sensitivity analyses using priors informed by observed cure rates in other randomised groups or strata will be performed and will be provided to the DMC to help inform their decision to stop a group.

**Sample size**

For the monitoring of single groups, assuming a target cure rate of 90%, an unacceptably true low cure rate of 70%, 90% power and one-sided alpha=0.05, and 5% loss to follow up by EOT+12, 39 patients would be required per group to exclude the null hypothesis that the cure rate was 90% based on a single-group test. There are 14 groups and 2 genotype strata, giving a total sample size of 1092 patients (39*14*2). For final comparisons at trial closure, the null hypothesis is that all groups will achieve the >90% cure target and be included in the final analysis. There will then be 546 per group for the regimen comparison, 156 in the control group and 312 per intervention group for the strategy comparison, and 468 per group for the ribavirin comparison.

Regulatory guidance recommends that non-inferiority margins be chosen to ensure that the difference between an intervention and the active control (here 12 weeks duration) would not exceed that between the active control group and a hypothetical placebo (or other standard control group, here the previous standard of care of 12-48 weeks PEG-IFN) (22). As cure rates with 12-48 weeks PEG-IFN were ~70% in genotype 6 (23), with similar or lower cure rates in other genotypes, and as we expect >90% SVR12 in all groups, both our non-inferiority margins of 5% for the regimen comparison and 10% for the strategy comparison would ensure this.

The choice of the non-inferiority margin was based on clinical judgement and the size of margins used in other trials of anti-infectives with relatively low failure rates such as community-acquired pneumonia (22). In particular, a smaller 5% margin is chosen for the drug comparisons because in practice they are likely to have similar advantages and disadvantages. In contrast, the different drug-sparing strategies have a variety of different advantages and disadvantages (in terms of additional visits vs. less drug vs. different drugs vs. weekends off, Table 1) which could be differentially balanced against overall cure rates, particularly considering impact on healthcare provision, e.g. through directly supervised therapy. Thus a greater non-inferiority margin is relevant to drug-shortening because the potential benefit in terms of numbers treated for the same fixed budget is much greater.

From initial power calculations, assuming that the overall cure rate is 95%, the numbers above provide 97% power to demonstrate non-inferiority between drug regimens based on a 5% margin and 96% power to demonstrate non-inferiority between shortening strategies based on a 10% margin, both with one-sided alpha=0.05. For superiority comparisons (conducted for ribavirin and any comparison that meets non-inferiority above) and two-sided alpha=0.05, these numbers provide >90% power to detect absolute differences in SVR12 of 5% for regimen or ribavirin comparisons and >80% power to detect absolute differences in SVR12 of 7% or more for the strategy comparisons.

**Statistical analysis**

The final analysis will estimate risk differences between groups using marginal effects after logistic regression. The model will include all main randomised effects and strata, and will test interactions between all randomisations. Interactions will only be included in the final model if the 95% credible interval for the interaction term excludes no effect (p<0.05 for frequentist analyses). The interaction between regimen and strategy will include all levels of strategy. The interaction between ribavirin and strategy will not include the standard treatment length strategy due to the partial factorial randomisation. Comparisons of regimens and of strategies will be non-inferiority analyses and the ribavirin comparison will be a superiority analysis. Primary analysis will be intention-to-treat using Bayesian methods and 90% credible intervals. Secondary analyses will consider per-protocol populations, frequentist methods, and 95% credible and confidence intervals.

Analysis priors for the final analysis are listed in Table 2; these differ from the monitoring priors as the aim of monitoring is only to identify poorly performing groups and not to compare the randomised groups. The control cure rate analysis prior is beta(4.75, 0.25), which has a mean of 0.95 and a variance derived as for the monitoring priors. Sensitivity analyses will use a range of informative priors reflecting plausible belief in the clinical community. Sceptical analysis prior distributions were chosen with means corresponding to the null hypothesis for each randomisation and enthusiastic analysis priors with means y greater than this, where y is the non-inferiority margin or absolute difference specified in the power calculations. The variances were arbitrarily set such that 90% of the prior distribution is within +/- y around the mean to reflect the strength of the belief in the mean effect. Thus for example, the risk difference for the drug regimen comparison has the sceptical analysis prior centred on -5% (the null hypothesis for the non-inferiority comparison) with 90% limits ±5%, giving a 0.05 probability that the cure rate will be 10% worse and a 0.05 probability the cure rate will increase (ie be >0%) (sceptical priors in each direction will be used for the regimen comparisons). The enthusiastic analysis prior is centred on 0% with 90% limits ±5%, giving a 0.05 probability that the cure rate will be 5% worse and a 0.05 probability that it will be 5% better with one regimen than the other.

To define the performance characteristics of the proposed stopping guideline, posterior probabilities of cure rates and the probability of stopping groups were calculated analytically using beta and binomial distributions respectively. Timings of interim analyses were determined by applying the probabilities of stopping groups to a projected recruitment schedule. The average probability of stopping a genuinely inferior group was estimated by integrating the probability of stopping a group with respect to the monitoring prior beta(4.5, 0.5) over cure rates between 60% and 90%. The lower bound was determined from previous studies testing strategies most similar to those in VIETNARMS which have reported cure rates of >90% with lower confidence interval bounds >60% (4, 5). In studies with cure rates <60%, all patients received shortened therapies, regardless of their HCV VL, and did not receive adjunctive drugs (24, 25), therefore they were not considered relevant to this analysis – although power would be even greater to stop such a group. Cure rates above 90% were not considered as groups with these cure rates should not be stopped and so do not affect timing of the analyses. Simulations of 5000 datasets with outcomes taken from binomial distributions were used to determine the cumulative probability of stopping groups at specified analysis time points, and to estimate power after being analysed using marginal effects after logistic regressions with a model containing all randomised comparisons, as described above. Predictive probabilities (the probability of achieving a success at the end of the trial) were calculated analytically using the beta-binomial distribution in R 3.5.1. All other analyses were performed using Stata v15.1.

**RESULTS AND DISCUSSION**

**Characteristics of Bayesian stopping guideline for individual groups**

The minimum number of failures required to satisfy the stopping criteria for the main monitoring beta(4.5, 0.5) prior and for each number of analysed patients are listed in Table 3. The probability of stopping a group is then the probability of observing the required number of failures in the group. When the true cure rate is 90%, the probability of incorrectly stopping a group is always <0.05, and this decreases as the true cure rate increases. It is expected from the specification of the stopping guideline that when the true cure rate is equal to the mean of the prior, then the probability of stopping a group is 0.05 so the probability of incorrectly stopping a group is always maintained below the correct level. The calculated probability is not exactly 0.05 and differs depending on the number of patients analysed due to the discrete nature of the outcome.

For small numbers of analysed patients, a larger proportion of failures are required to stop a group, increasing from a minimum of 17% to 100% of those analysed; therefore the probability of stopping a group incorrectly early in recruitment is also smaller as there is a smaller chance of observing greater proportions of failures regardless of the true cure rate. This protects against groups being stopped erroneously due to a high concentration of failures amongst the initial patients reaching EOT+12, and if the two strata within a group share the same true cure rate then it is unlikely only one will reach the stopping threshold.

Groups with the lowest cure rates considered plausible (60%) are highly likely to reach the stopping criteria quickly (>90% chance of stopping after analysing 21-26 patients). Groups with moderately low cure rates (<80%) are also likely to be stopped before recruitment ends. However, groups with true cure rates slightly under 90% are unlikely to be stopped (results not shown). A low chance of stopping a group just below the target cure rate might be considered unacceptable in futility stopping guidelines in other situations, but the target of 90% is largely arbitrary and there may be interest in strategies that have a slightly lower cure rate if they are able to expand treatment access to difficult to reach populations. Increasing the probability of stopping groups with cure rates just below 90% would lead to a greater chance of incorrectly stopping groups with cure rates >90% and it is considered more important to retain these than to stop groups with slightly lower cure rates. Additionally, any other stopping guideline would similarly be unable to discriminate between these cure rates without a very large sample size. The probability of incorrectly not stopping a group rapidly decreases as the true cure rate decreases.

**Timing of interim analyses**

Interim analyses need sufficient numbers of patients at EOT+12 to give a reasonable probability of stopping a genuinely inferior group. It was therefore decided to perform analyses after the first month such that at least one inferior group has a 0.3, 0.5 or 0.7 average probability of being stopped were this to be the first interim analysis, assuming cure rates are uniformly distributed on [0.6, 0.9] and given projected recruitment. An average probability is used to reflect the uncertainty about the true cure rates; for low cure rates the probability of stopping a group can be substantially higher (Figure 2). An additional analysis before these thresholds will allow for any groups with a very low cure rate below the anticipated minimum of 60% to be detected early despite the very small probability of detecting a group with a cure rate between 60-90%.

Four analysis timepoints were chosen to provide multiple opportunities to detect failing groups while allowing adequate time between analyses for the accrual of patients and outcome data, and preventing unnecessary burden on time and resources needed for analyses and subsequent DMC meetings. The highest probability threshold (0.7) is determined by the maximum average cure rate of the genuinely inferior groups and by the recruitment schedule, as analyses need to be performed sufficiently early enough to gain the benefit of randomising remaining patients to the other groups. The other thresholds (0.3, 0.5) were evenly spaced across the probabilities of stopping a group with an average cure rate, taking into consideration the first, early DMC meeting not based on these probabilities.

Based on the underlying projected recruitment, we therefore expect the interim analyses to take place after 7, 10, 13 and 18 months. The number of patients in each group and the probability of stopping an inferior group of each strategy type at these analyses are listed in Table 4. By assessing the cumulative probability of stopping a group (Supplementary Figure 1, Additional File 1), our schedule provides a balance between having more frequent meetings, which have a lower probability of stopping groups at any individual meeting but allow for earlier detection of poorly performing groups, and less frequent meetings, which have a higher probability of stopping a group but its low performance is detected later. The exception to this is having analyses every month, but this schedule is impractical due to the resources required for an interim analysis.

As this is only projected recruitment, sensitivity analyses were performed to examine the effect of faster or slower recruitment (Supplementary Table 1, Additional File 1). These indicate that changes to the recruitment schedule alter only the timing of the analyses; the number of patients in each analysis differs by less than the estimated number recruited in 1 month. Therefore, if there are significant delays in recruitment, interim analyses will be timed such that they include a similar number of patients at EOT+12 to that in the expected schedule. Sensitivity analyses also explored changing the lower bound of the distribution over which cure rates of genuinely inferior groups are assumed to be distributed to below 60% (Supplementary Table 2, Additional File 1), but this had minimal effect on the timing of initial interim analyses. There were greater differences in the timings of the last analyses, but the timing of this analysis is the most flexible and can be determined based on observed rather than assumed true cure rates.

**Power for final analysis**

Given the lack of knowledge of the standard 12-week cure rates for this population (anticipated 50% of patients with genotype 6) and how cure rates in the shortened treatment with ribavirin groups will compare to these and to shortened treatment without ribavirin, there is uncertainty regarding the overall power for the final analysis even if the overall cure rate is 95% (Table 5). If we assume equality between regimens and a 5% absolute difference for ribavirin, then these constraints mean that the cure rates in each group are completely determined by the difference between the shortening strategies. Non-inferiority can exist between the standard duration group and either the pooled shortening strategy groups, the shortening strategy without ribavirin groups, or the shortening strategy with ribavirin groups, meaning the shortening strategy with ribavirin groups can have cure rates 2.5% higher, 5% higher or equal to the standard duration groups respectively. These alternatives are shown in different columns of Table 5.

Power to determine non-inferiority for the regimen comparison using a 5% margin is mostly unaffected by assumptions about different values for the standard 12-week cure rate and effect of ribavirin, with power remaining close to 100%. Power to determine non-inferiority in the strategy comparison using a 10% margin is similarly unaffected, with power remaining close to 100% when comparing the pooled shortening strategy groups against standard duration. For superiority comparisons, power to detect a 5% absolute difference in the ribavirin or regimen comparison remains high at >90% regardless of the standard duration cure rate and ribavirin effect.

**Limitations of the design**

A potential weakness in the design is that the sample size was not originally calculated using Bayesian principles, but primary analyses will be conducted using Bayesian methods to allow for the calculation of posterior probabilities exploring the difference in cure rates between the interventions. However, for the non-inferiority comparisons, sample size estimates obtained using Bayesian methods are similar to or smaller than those obtained using frequentist methods (26), suggesting that our design is likely to be conservative. Additionally, secondary analyses will use frequentist methods for comparison. For interim analyses, the probability of correctly stopping a group, analogous to the frequentist concept of power, is determined by the true cure rate in the group and the number of analysed patients at each analysis, not by the overall group size.

The timing of and the number of patients at interim analyses are determined by at least one group, usually the 4 week treatment group with PEG-IFN since this has the shortest overall treatment duration, reaching a certain probability threshold of being stopped. This may mean delays in identifying unsuccessful groups receiving other strategies. This is unlikely if groups have cure rates lower than the average cure rate because, as discussed above, these will be detected faster than anticipated, but delays may occur if the cure rate is above the average, but <90%. As the treatment length of patients in the RGT groups is unknown until after their day 7 visit, it is not possible to stagger treatment start dates so that the length between randomisation and EOT is the same for all strategies. Staggered treatment start might also lead to drop out after randomisation but before starting treatment, leading to inefficiency and potential bias. Cure rates will be monitored and if the timing rule is found to be inappropriate it can be adjusted to change the timing or frequency of analyses with no penalty, due to the use of Bayesian monitoring (16).

The power calculations for the final analysis assume that all groups will be included and that no groups have been stopped. It is possible that power will be lower if fewer groups are included, but for most comparisons with a full sample, power is very high and is likely to remain acceptable at the final analysis even with the exclusion of some patients. To help preserve power, if groups are stopped early subsequent patients will be randomised to open groups. The power calculations were also estimated using frequentist methods, though the primary analysis will use Bayesian methods. However as power is extremely high, the analogous concept to power in Bayesian analysis, that for non-inferiority comparisons the lower credible interval bound is above the non-inferiority margin, is likely to be similarly high.

**Alternative designs**

Alternative Bayesian designs include basing the stopping guideline on a predictive probability, the probability of achieving a success at the end of the trial. In VIETNARMS, a success during the monitoring period is stopping a genuinely inferior group, which means that there is a >0.95 posterior probability of a <90% true cure rate in that group. A rule based on predictive probabilities would then state a group will be stopped at an interim analysis if there is a >0.95 chance of stopping a group at the end of the trial (Pr([Pr(true cure rate<0.9|z)>0.95]|x)>0.95 where x is the data currently observed and z the complete data with all outcomes observed). For the monitoring beta(4.5, 0.5) prior and a fully recruited group, the stopping criteria is met with 13 failures so equivalently the group is stopped if there is a >0.95 chance of observing ≥13 failures in the fully recruited group.

Predictive probabilities place a large emphasis on the arbitrary target cure rate of 90%, and hence was not used for VIETNARMS. The final analysis will compare strategies against control and not test cure rates in individual groups. The aim of monitoring is to detect poorly performing groups and stop them early, rather than to ultimately meet a particular cure rate within a group at the end of the trial, as there may be other advantages to strategies that have a slightly lower cure rate than 90% in specific populations or circumstances. Compared to the posterior probability based stopping guideline, using predictive probabilities requires a similar number of failures or more to stop a group (Supplementary Table 3, Additional File 1) so they do not offer any benefits in detecting poorly performing groups more quickly for our design. Stopping rules and guidelines based on posterior probabilities can be converted to those based on predictive probabilities (27) so interim analyses can incorporate predictive probabilities to provide more information to the DMC.

Another approach would be to analyse the outcome data after every reported outcome rather than at scheduled interim analyses, which could reduce the time until a genuinely inferior group is stopped. Implementing this would be complex due to the many groups and varying treatment lengths. The small benefit in the reduction in time would not justify the additional work required to monitor outcomes intensely.

**CONCLUSIONS**

We have designed a trial allowing for the testing of multiple approaches to drug choice and shortening strategy for HCV treatment. We have closely examined the statistical aspects of the trial, with particular focus on the implications of the chosen rule for early stopping of unsuccessful groups. We have shown that the operating characteristics of the rule are appropriate and that interim analyses can be timed to detect failing groups at various stages.

Given the pressures on funding and time, it is desirable to test many aspects of treatment at once and to allow for the swift removal of unsuccessful strategies: Bayesian monitoring methods allow for this. Despite the focus on HCV treatment, the statistical principles behind our novel design are not limited to this area and could be applied to other clinical trials in a wide variety of settings.

**Abbreviations**

DAA: direct acting antivirals; DMC: data monitoring committee; EOT+12: 12 weeks after the end of treatment; ESS: estimated sample size; HCV: hepatitis C virus; LLOQ: lower limit of quantification;

PEG-IFN: pegylated interferon; SVR12: sustained virologic response 12 weeks after the end of treatment; VL: viral load

**DECLARATIONS**

**Ethics approval and consent to participate**

Not applicable

**Consent for publication**

Not applicable

**Availability of data and materials**

Not applicable

**Competing interests**

The authors declare that they have no competing interests.

**Funding**

VIETNARMS is funded by the Wellcome Trust (206296/Z/17/Z). GSC is supported in part by the BRC of Imperial College NHS Trust and an NIHR Research Professorship. ASW is an NIHR Senior Investigator. EB was funded by the Medical Research Council UK, the Oxford NIHR Biomedical Research Centre and is an NIHR Senior Investigator. The views expressed are those of the author(s) and not necessarily those of the NHS, the NIHR or the Department of Health.

**Authors' contributions**

LM performed the statistical analysis and wrote the first draft of the manuscript. GSC, NVVC, EB, SLP and ASW designed the trial. IRW, GSC and ASW provided critical comments on initial drafts of the manuscript. All authors read and approved the manuscript.

**Acknowledgements**

The SEARCH consortium would like to acknowledge STOP-HCV (MR/K01532X/1), supporting stratified medicine studies in HCV.

SEARCH Investigators (alphabetical order): Eleanor Barnes, Graham S Cooke, Jeremy N Day, Nguyen Thanh Dung, Barnaby Flower, Tim Hallett, Le Manh Hung, Evelyne Kestelyn, Dao Bach Khoa, Leanne McCabe, Sarah L Pett, Le Thanh Phuong, Motiur Rahman, Joel Tarning, Hugo C Turner, Guy E Thwaites, Nguyen Van Vinh Chau, A Sarah Walker, Nicholas J White.

**Tables and figures**

**Table 1: Components of each shortening strategy and cost differences compared with standard 12-week treatment**

| Shortening strategy | Resource use of providing treatment | | | | Difference in cost compared to standard 12-week treatment |
|---|---|---|---|---|---|
| | Mean weeks of DAA | Number of PEG-IFN injections | Number of visits required | Number of VL tests | |
| Standard 12-week treatment | 12 | 0 | 3* | 1* | - |
| PEG-IFN+DAA (4 weeks treatment, weekly PEG-IFN) | 4 | 4 | 6 | 1 | -8x(weekly drug cost) + 4x(interferon cost) + 3x(visit cost) |
| Response guided therapy (4, 8, or 12 weeks treatment) | 9.6** | 0 | 4 | 2 | -3.4x(weekly drug cost) + 1x(visit cost) +1x(VL cost) |
| Induction/maintenance (7 days/week for 2 weeks, then 5 days/week) | 9.14 | 0 | 4† | 1 | -3.86x(weekly drug cost) + 1x(visit cost) |

\* assuming minimum visits at treatment initiation, end of treatment, end of treatment plus 12 weeks (where VL is done once to assess cure)

\*\* assuming 1:3:1 receiving 4:8:12 weeks treatment. One extra visit and VL at day 7 to assess initial VL response

† assuming one extra visit at week 2 when move from induction to maintenance phase

PEG-IFN: pegylated-inteferon; DAA: direct acting antiviral.

**Table 2: Priors to be used in final analysis**

| | Primary analysis: uninformative | Sensitivity analyses | |
|---|---|---|---|
| | | Sceptical | Enthusiastic |
| Control cure rate | Beta(4.75, 0.25) | Beta(4.75, 0.25) | Beta(4.75, 0.25) |
| Regimen comparison | N(0, 10000) | N(-0.05, 0.009) | N(0, 0.009) |
| Strategy comparison | N(0, 10000) | N(-0.1, 0.0036) | N(0, 0.0036) |
| Ribavirin comparison | N(0, 10000) | N(0, 0.009) | N(0.05, 0.009) |

The beta prior for the control cure rate has mean 0.95 and variance 0.008.

The model will also be adjusted for genotype, which will have the prior N(0, 10000) in all analyses.

N(m,v) is the normal prior with mean m and variance v.

**Table 3: Number of failures needed to stop a drug strategy group for each number of analysed patients**

| Analysed patients | Minimum number of observed failures needed to stop group* | Maximum probability of stopping group if true cure rate equals: | | Minimum probability of stopping group if true cure rate equals: | | | |
|---|---|---|---|---|---|---|---|
| | | 90% | 95% | 90% | 80% | 70% | 60% |
| 3-7 | 3 | 0.026 | 0.004 | 0.001 | 0.008 | 0.027 | 0.064 |
| 8-13 | 4 | 0.034 | 0.003 | 0.005 | 0.056 | 0.194 | 0.406 |
| 14-20 | 5 | 0.043 | 0.003 | 0.009 | 0.130 | 0.416 | 0.721 |
| 21-26 | 6 | 0.040 | 0.002 | 0.014 | 0.231 | 0.637 | 0.904 |
| 27-33 | 7 | 0.042 | 0.001 | 0.015 | 0.287 | 0.744 | 0.958 |
| 34-39** | 8 | 0.037 | 0.001 | 0.017 | 0.367 | 0.844 | 0.986 |
| 40-41** | 8 | 0.048 | 0.001 | 0.042 | 0.563 | 0.945 | 0.998 |
| 42-48 | 9 | 0.046 | 0.001 | 0.021 | 0.469 | 0.920 | 0.997 |
| 49-55 | 10 | 0.044 | 0.0004 | 0.022 | 0.528 | 0.952 | 0.999 |
| 56-63 | 11 | 0.047 | 0.0003 | 0.021 | 0.580 | 0.971 | 1.000 |
| 64-71 | 12 | 0.048 | 0.0002 | 0.023 | 0.648 | 0.985 | 1.000 |
| 72-78 | 13 | 0.045 | 0.0001 | 0.025 | 0.705 | 0.993 | 1.000 |

* >0.95 posterior probability of the true cure rate being <90% (Pr(true cure rate <0.9|x)>0.95, where x is the data currently observed, with the beta(4.5, 0.5) prior which has mean=0.9 and variance=0.015

**These rows not pooled despite the same number of minimum failures to provide information about the fully recruited strata (n=39).

Note: Groups will recruit 78 patients with 39 patients in each stratum.

Maximum and minimum probabilities are for the range of analysed patients.

**Table 4: Timing of interim analyses**

|  | First | Second | Third | Fourth |
|---|---|---|---|---|
| Months since recruitment started | 7 | 10 | 13 | 18 |
| Total recruited | 205 | 361 | 517 | 777 |
| Total at EOT+12 weeks | 44 | 144 | 286 | 533 |
| At EOT+12 weeks in each: |  |  |  |  |
|   PEG-IFN group | 5 | 14 | 24 | 42 |
|   RGT group | 3 | 11 | 21 | 39 |
|   Induction/maintenance group | 2 | 8 | 17 | 35 |
| Average probability genuinely inferior group will be stopped*: |  |  |  |  |
|   PEG-IFN groups | 0.124 | 0.297 | 0.537 | 0.710 |
|   RGT groups | 0.021 | 0.313 | 0.440 | 0.665 |
|   Induction/maintenance groups | 0.070 | 0.150 | 0.431 | 0.593 |

*Assuming true cure rate uniformly distributed over 60-90%

EOT+12: 12 weeks after the end of treatment; PEG-IFN: pegylated-interferon; RGT: response guided therapy

**Table 5: Group-specific cure rates and power for 1092 patients with an overall 95% cure rate**

|  | Ribavirin group cure rate compared to 12 week standard treatment cure rate | | |
|---|---|---|---|
|  | 5% higher | 2.5% higher | Equal |
| Group-specific cure rates: |  |  |  |
|   Standard 12 week treatment groups | 93.3% | 95% | 96.7% |
|   Shortened treatment with ribavirin groups | 98.3% | 97.5% | 96.7% |
|   Shortened treatment without ribavirin groups | 93.3% | 92.5% | 91.7% |
| Power for a: |  |  |  |
|   5% non-inferiority margin for regimen comparison | 99% | 98% | 97% |
|   10% non-inferiority margin for strategy comparison | 100% | 100% | 96% |
|   5% absolute difference for ribavirin comparison | 98% | 95% | 91% |

Note: Group-specific cure rates are such that overall cure rate in 1092 patients is 95%, ribavirin effect between shortened treatment groups is 5% and there is no effect of regimen. The effect of ribavirin in shortened treatment compared to standard treatment is varied to reflect uncertainties in the cure rates and to vary the no effect of strategy between standard treatment and shortened treatment without ribavirin, shortened treatment with ribavirin and pooled shortened treatment groups. Power is calculated with one sided alpha=0.05 for non-inferiority margins and two-sided alpha=0.05 for superiority comparisons reflecting the design.

**Figure Legends**

**Figure 1: Trial Schema**

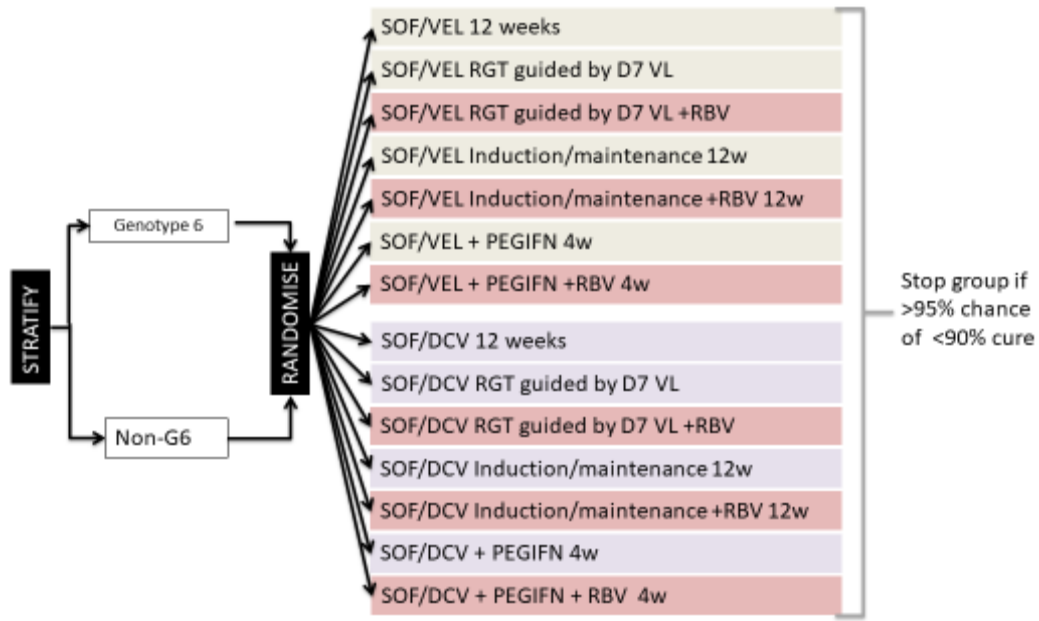

Note: SOF/VEL=sofosbuvir/velpatasvir, SOF/DCV=sofosbuvir/daclatasvir, RBV=ribavirin, VL=viral load, RGT=response guided therapy, PEGIFN=pegylated interferon.

**Figure 2: Initial probability of stopping groups over an estimated recruitment schedule for various true cure rates**

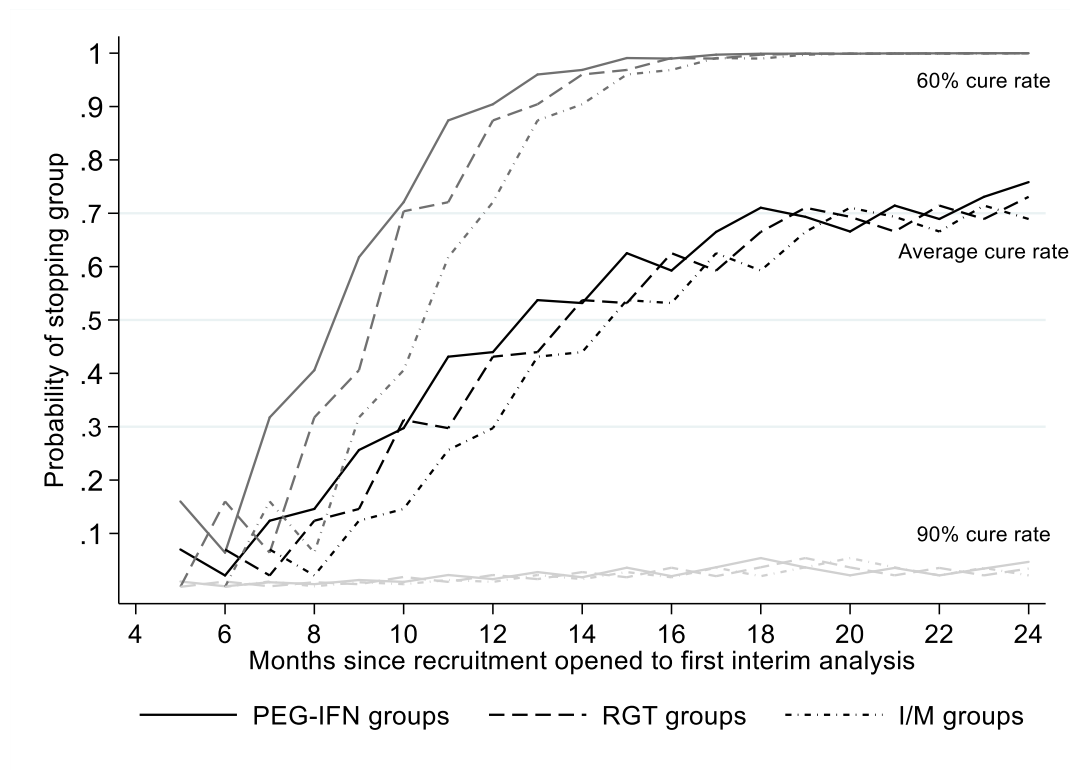

Note: The average true cure rate is uniformly distributed over 60-90%. Probabilities are calculated assuming no previous interim analysis. The total number of patients recruited at each month, of a total target of 1092, is: 113 at 5 months, 153 at 6, 205 at 7, 257 at 8, 309 at 9, 361 at 10, 413 at 11, 465 at 12, 517 at 13, 569 at 14, 621 at 15, 673 at 16, 725 at 17, 777 at 18, 829 at 19, 881 at 20, 933 at 21, 985 at 22, 1037 at 23 and 1092 at 24. See Supplementary Figure 1, Additional File 1 for the cumulative probability of stopping a group.